\documentclass[fleqn,twoside,twocolumn,nofootinbib,showkeys]{revtex4} 
\usepackage[sec,nocpr,nopacs]{ujp} 

\newcommand{\ts}{\textstyle}

\begin{document}
\title[Generalized Equidistant Che\-by\-shev Polynomials  ]
{GENERALIZED EQUIDISTANT CHEBYSHEV POLYNOMIALS AND ALEXANDER KNOT INVARIANTS\,$^{{}^{\mbox{\footnotesize1}}}$}%
\author{A.M.~Pavlyuk}
\affiliation{Bogolyubov Institute for Theoretical Physics, Nat. Acad. of Sci. of Ukraine}
\address{14b, Metrolohichna Str., Kyiv 03143, Ukraine}
\email{pavlyuk@bitp.kiev.ua}
\udk{539.12; 517.984}  \razd{\seciii}

\autorcol{A.M.\hspace*{0.7mm}Pavlyuk}%

\setcounter{page}{488}%

\begin{abstract}
We introduce the generalized equidistant Che\-by\-shev polynomials
$T^{(k,h)}$ of kind $k$ of hyperkind $h,$ where $k,h$ are positive
integers.\,\,They are obtained by a generalization of standard and
monic Che\-by\-shev  polynomials of the first and second
kinds.\,\,This generalization is fulfilled in two directions.\,\,The
horizontal generalization is made by introducing hyperkind $h$ and
expanding it to infinity.\,\,The vertical generalization proposes
expanding kind $k$ to infinity with the help of the method of
equidistant coefficients.\,\,Some connections of these polynomials
with the Alexander knot and link polynomial invariants are
investigated.
\end{abstract}

\maketitle

\section{Introduction}\vspace*{-1.5mm}

Knots and links appear in many areas of physics, par-\linebreak
ti\-cu\-lary in electrodynamics, optics, liquid crystals,
hy\-dro\-dy\-na\-mics \cite{At,Ka,Ka2,RV}.\,\,The main
characteristics of knots
 and links follow from the axioms of skein
relation and nor\-ma\-li\-za\-tion, which lead to polynomial
in\-va\-riants \cite{Co}.\,\,Each knot and link is described by some
de\-fi\-nite po\-ly\-no\-mial.\,\,The most known polynomial
in\-va\-riants are Alexander \cite{Al}, Jones~\cite{Jo} and
\mbox{HOMFLY} \cite{HO} ones.

The Che\-by\-shev polynomials appear in different bran\-ches of
mathematics, particulary in app\-ro\-xi\-ma\-tion theory, knot
theory, combinatorics, dif\-fe\-ren\-tial equations, number theory
\cite{Ri}.\,\,As important ma\-the\-ma\-ti\-cal inst\-ru\-ments,
they  widely penetrate into dif\-fe\-rent branches of
phy\-sics.\,\,We exploit the Che\-by\-shev polynomials and their
ge\-ne\-ra\-li\-za\-tions for the description of knots and links
\cite{GP1,GP2,Pa1}.\,\,To our mind, they have to play important role
even as in the searching of new  polynomial invariants.

In the present paper, we introduce the generalized equidistant
Chebyshev polynomials $T^{(k,h)}$ of kind $k$ of hyperkind $h,$
where $k,h$ are positive integers.\footnotetext[1]{This paper was
presented at the 3rd Walter Thirring International School on
Fundamentals of Astroparticle and  Quantum Physics (17--23
September, 2017, BITP, Kyiv,\linebreak Ukraine).}\,\,At the
beginning, we unify the standard Chebyshev polynomials of the first
kind  and monic Chebyshev polynomials of the first kind with the
help of one formula using the newly introduced notion of hyperkind
$h$ ($h=1$ refers to all standard Chebyshev polynomials; $h=2$~-- to
all monic Chebyshev polynomials).\,\,After that, a generalization
means declaring $h$ to be any positive integer and leads to
generalized Chebyshev polynomials of the first
kind.\,\,Ana\-lo\-gous\-ly, one obtains generalized Che\-by\-shev
polynomials of the second kind.\,\,Thus, we fulfil the so-called
horizontal generalization.\,\,The final step is making the vertical
generalization, by using the method of equidistant
coefficients.\,\,As a result, we expand the meanings of kind $k$ to
all positive integers and obtain the formula for generalized
equidistant Chebyshev polynomials of kind $k$ of
hyperkind~$h$.\looseness=1

\section{Standard Che\-by\-shev Polynomials}

The standard Che\-by\-shev polynomials of the {\it first kind ${
T}^{(1)}_{n}(x)$} are introduced as\vspace*{-1mm}
  \begin{equation}
  \label{sT}{ T}^{(1)}_{n}(x)=\cos
(n\theta),\quad  x=\cos\theta,
  \end{equation}
where $n$  is a nonnegative integer.\,\,In our paper, the
Che\-by\-shev polynomials are characterized by two numbers~--
$(k,h).$  Here, the letter $k$  stands for {\it ``kind''}.\,\,In
addition, we introduce a new notion of {\it ``hyperkind''}, for
which we use the upper index $h$ (in brackets); $h=1$ refers to all
standard Che\-by\-shev polynomials.\,\,Thus, for all polynomials
defined by~(\ref{sT}), $(k,h)=(1,1)$.\,\,In the case of monic
Che\-by\-shev polynomials considered in the next section,
$h=2$.\,\,The relation
\begin{equation}\label{coscos} \cos{(n+1)\theta}+\cos{(n-1)\theta}=2\cos{\theta}\cos{n\theta}
\end{equation}
yields the recurrence relation for the standard Che\-by\-shev
polynomials of the first kind:\vspace*{-1mm}
 \begin{equation}\label{rec-sT} {T}^{(1)}_{n+1}(x)=2x{T}^{(1)}_{n}(x)-{ T}^{(1)}_{n-1}(x),~ {
 T}^{(1)}_{0}=1,~
{ T}^{(1)}_{1}=x,
\end{equation} which can be considered as
an alternative definition of $T^{(1)}_{n}(x)$.\,\,Some examples of
${ T}^{(1)}_{n}(x)$ obtained from~(\ref{rec-sT}) are as
follows:\vspace*{-1mm}
 \begin{equation}\label{sTx}
 \begin{array}{l}
{T}^{(1)}_{0}=1,~ { T}^{(1)}_{1}=x,~
   {T}^{(1)}_{2}=2x^{2}-1,\\[1mm]
{T}^{(1)}_{3}=4x^{3}-3x,~ {T}^{(1)}_{4}=8x^{4}-8x^{2}+1,\\[1mm]
{T}^{(1)}_{5}=16x^{5}-20x^{3}+5x.
 \end{array}
 \end{equation}

The standard Che\-by\-shev polynomials of the {\it second kind} ${
V}^{(1)}_{n}(x)$ are defined  in the following way  $(k=2$,
$h=1$):\vspace*{-3mm}
\begin{equation}\label{sV}{ V}^{(1)}_{n}(x)={\sin(n+1)\theta\over {\sin\theta}},\quad
x=\cos\theta .
  \end{equation}
  The relation
\begin{equation}\vspace*{-1mm}
\label{sinsin}
\sin{(n+2)\theta}+\sin{n\theta}=2\cos{\theta}\sin{(n+1)\theta}
\end{equation}
implies that definition~(\ref{sV})
 can be written in terms of the recurrence relation similar to~(\ref{rec-sT}):\vspace*{-1mm}
 \begin{equation}\label{rec-sV} { V}^{(1)}_{n+1}(x)=2x{ V}^{(1)}_{n}(x)-{ V}^{(1)}_{n-1}(x),\ { V}^{(1)}_{0}=1,
{V}^{(1)}_{1}=2x.
 \end{equation}
Some  first standard Che\-by\-shev polynomials of the second kind
are:
 \begin{equation}\label{sVx}
 \begin{array}{l}
{V}^{(1)}_{0} = 1,~ { V}^{(1)}_{1}\!= \!2x, ~{V}^{(1)}_{2} = 4x^{2}-1,\\[1mm]
{V}^{(1)}_{3} = 8x^{3}-4x,~
{V}^{(1)}_{4}=16x^{4}-12x^{2}+1,\\[1mm]
{V}^{(1)}_{5}=32x^{5}-32x^{3}+6x.
 \end{array}\!\!\!\!\!\!\!\!\!\!\!\!\!\!\!\!
  \end{equation}
 From the formula
\begin{equation}
\label{2cos}2\cos(n\theta)=\Bigl(\!{\sin(n+1)\theta\over
{\sin\theta}}-{\sin(n-1)\theta\over {\sin\theta}} \!\Bigl)\!,
  \end{equation}
one finds that definition~(\ref{sT}) of $T^{(1)}_{n}(x)$ can be
rewritten in the form
\begin{equation}\label{sT-} T^{(1)}_{n}(x)={1\over 2}{\sin(n+1)\theta\over {\sin\theta}}-{1\over 2}{\sin(n-1)\theta\over {\sin\theta}},
\quad  x=\cos\theta .
  \end{equation}

As follows from~(\ref{2cos}), the Che\-by\-shev polynomials
$T^{(1)}_{n}(x)$ and $V^{(1)}_{n}(x)$ are connected by the formula
 \begin{equation}\label{con1}
2T^{(1)}_{n}(x)=V^{(1)}_{n}(x)-V^{(1)}_{n-2}(x) , \quad n\ge
0.
\end{equation}
Here, we mean that $V^{(1)}_{-1}=0$, $ V^{(1)}_{-2}=-1$, which
follows from~(\ref{rec-sV}).

\section{Monic Che\-by\-shev Polynomials}

The monic Che\-by\-shev polynomials of the   
{\it first kind ${ T}^{(2)}_{n}(x)$}, $(k=1$, $ h=2)$, are defined
by
  \begin{equation}\label{mT} T^{(2)}_{n}(x)=2\cos
(n\theta),\quad x=2\cos\theta.
  \end{equation}
The monic normalization means that all Che\-by\-shev polynomials
have the unit coefficients at the highest degree $x^{n}$.

 The recurrence relation
 \begin{equation}\label{rec-mT} T^{(2)}_{n+1}(x)=xT^{(2)}_{n}(x)-T^{(2)}_{n-1}(x),\ T^{(2)}_{0}=2,\
 T^{(2)}_{1}=x,
 \end{equation}
 can be considered as an alternative definition of $T^{(2)}_{n}(x)$.\,\,Some first cases of $T^{(2)}_{n}(x)$ are:
 \begin{equation}\label{mTx}
 \begin{array}{l}
T^{(2)}_{0}=2,~ T^{(2)}_{1}=x,~
   T^{(2)}_{2}=x^{2}-2,\\[1mm]
T^{(2)}_{3}=x^{3}-3x,~ T^{(2)}_{4}=x^{4}-4x^{2}+2,\\[1mm]
T^{(2)}_{5}=x^{5}-5x^{3}+5x.
 \end{array}
  \end{equation}

The monic Che\-by\-shev polynomials of the 
{\it second kind ${ V}^{(2)}_{n}(x)$} are defined  as $(k=2$,
$h=2)$
  \begin{equation}\label{mV} V^{(2)}_{n}(x)={\sin(n+1)\theta\over {\sin\theta}},\quad
x=2\cos\theta
  \end{equation}
or, in terms of the recurrence relation,
 \begin{equation}\label{rec-mV} V^{(2)}_{n+1}(x)=xV^{(2)}_{n}(x)-V^{(2)}_{n-1}(x),\ V^{(2)}_{0}=1,\
V^{(2)}_{1}=x.
 \end{equation}
  The first monic Che\-by\-shev polynomials of
the second kind are:
 \begin{equation}\label{mVx}
 \begin{array}{l}
V^{(2)}_{0}=1,~ V^{(2)}_{1}=x,~ V^{(2)}_{2}=x^{2}-1,\\[1mm]
V^{(2)}_{3}=x^{3}-2x,~ V^{(2)}_{4}=x^{4}-3x^{2}+1,\\[1mm]
V^{(2)}_{5}=x^{5}-4x^{3}+3x.
 \end{array}
  \end{equation}

Definition~(\ref{mT}) of $\,T^{(2)}_{n}(x)$ can be written in the
form\vspace*{-3mm}
\begin{equation}\label{mT-} T^{(2)}_{n}(x)={\sin(n\!+\!1)\theta\over {\sin\theta}}-{\sin(n\!-\!1)\theta\over {\sin\theta}},~   x=2\cos\theta .
  \end{equation}

The Che\-by\-shev polynomials $T^{(2)}_{n}$ and $V^{(2)}_{n}$ are
connected by the formula
 \begin{equation}\label{con1}
 T^{(2)}_{n}=V^{(2)}_{n}-V^{(2)}_{n-2} , \quad  n\ge 0,
\end{equation}
where $V^{(2)}_{-1}=0$, $V^{(2)}_{-2}=-1$, being calculated
from~(\ref{rec-mV}).

\section{Horizontal Generalization\\ of  Che\-by\-shev  Polynomials: Expanding\\ Hyperkind \boldmath$h$  to Infinity}

The main idea for the horizontal generalization of Che\-by\-shev
polynomials consists of two steps.\,\,The first step is a
unification of standard Che\-by\-shev polynomials and monic
Che\-by\-shev polynomials of the same kinds.\,\,At the beginning, we
unify ($k=1$, $h=1$) and ($k=1$, $h=2$) Che\-by\-shev polynomials
with the help of one formula; another formula unifies ($k=2$, $h=1$)
and ($k=2$, $h=2$) Che\-by\-shev polynomials.\,\,These formulas
automatically lead to the second step of
the horizontal generalization, by expanding the meanings of  hyperkind $h$ to infinity.\,\,

\subsection{Generalized Che\-by\-shev\\ polynomials of the first kind}

Thus, comparing~(\ref{sT}) and~(\ref{mT}), we write down the unified
Che\-by\-shev polynomials of the first kind ${ T}^{(h)}_{n}(x)$  by
the formula:
  \begin{equation}\label{gT}{ T}^{(h)}_{n}(x)=h\cos
(n\theta),\quad x=h\cos\theta,         
  \end{equation}
 for $h=1$ and $h=2$.
From this formula, one obtains the generalized Che\-by\-shev
polynomials of the first kind, by expanding $h$ to infinity, i.e.,
$h$ being any positive integer: $\ h=1,2,3,4,...\,$.

Relation (\ref{coscos}) written in the form\vspace*{-2mm}
\[
h\cos{(n+1)\theta}+h\cos{(n-1)\theta}={2\over h}\, h\cos{\theta} \,
h\cos{n\theta}
\]\vspace*{-5mm}

\noindent yields
 the recurrence relation for the generalized Che\-by\-shev polynomials of the first kind\vspace*{-1mm}
 \begin{equation}
 \label{rec-gT}\begin{array}{l} \displaystyle{T}^{(h)}_{n+1}(x)={2\over h}x{T}^{(h)}_{n}(x)-{
 T}^{(h)}_{n-1}(x),\\[3mm]
 { T}^{(h)}_{0}=h,~
{ T}^{(h)}_{1}=x, \end{array}
\end{equation}\vspace*{-3mm}

\noindent which can be taken for the definition of $T^{(h)}_{n}(x)$.

Some examples of ${ T}^{(h)}_{n}(x)$ obtained from~(\ref{rec-gT})
are:\vspace*{-1mm}
\begin{equation} \label{gTx}
 \begin{array}{l}
\displaystyle{T}^{(h)}_{0}=h,~ { T}^{(h)}_{1}=x,~
   {T}^{(h)}_{2}={2\over h}x^{2}-h, \\[3mm]
\displaystyle{T}^{(h)}_{3}={4\over h^2}x^{3}-3x,~ {T}^{(h)}_{4}={8\over h^3}x^{4}-{8\over h}x^{2}+h, \\[3mm]
\displaystyle{T}^{(h)}_{5}={16\over h^4}x^{5}-{20\over h^2}x^{3}+5x.
 \end{array} \end{equation}\vspace*{-3mm}

 \noindent
From~(\ref{gT}),~(\ref{rec-gT}), and~(\ref{gTx}), one has formulas
for the standard Che\-by\-shev polynomials of the first kind, if
$h=1$, and for the monic Che\-by\-shev polynomials of the first kind
in the case $h=2$.

As an example, let us write (generalized) Che\-by\-shev polynomials
of the first kind of the third hyperkind. For $(k= 1$, $ h=3),$ one
has\vspace*{-1mm}
\begin{equation}\label{3T}{ T}^{(3)}_{n}(x)=3\cos
(n\theta),~ 3\cos\theta=x,
  \end{equation}\vspace*{-9mm}
  \begin{equation}\label{rec-3T} {T}^{(3)}_{n+1}={2\over 3}x{T}^{(3)}_{n}-{ T}^{(3)}_{n-1},~ {
  T}^{(3)}_{0}=3,~
{ T}^{(3)}_{1}=x,
\end{equation}\vspace*{-9mm}
\begin{equation} \label{3Tx}
 \begin{array}{l}
\displaystyle{T}^{(3)}_{0}=3,~ { T}^{(3)}_{1}=x,~
   {T}^{(3)}_{2}={2\over 3}x^{2}-3,  \\[2mm]
\displaystyle{T}^{(3)}_{3}={4\over 9}x^{3}-3x,~
{T}^{(3)}_{4}={8\over 27}x^{4}-{8\over 3}x^{2}+3,  \\[3mm]
\displaystyle{T}^{(3)}_{5}={16\over 81}x^{5}-{20\over 9}x^{3}+5x.
 \end{array}
 \end{equation}

As follows from~(\ref{2cos}), definition~(\ref{gT}) of
$\,T^{(h)}_{n}(x)$ can be also written as\vspace*{-1mm}
\begin{equation}
\label{gT-} T^{(h)}_{n}(x)={h\over 2}\, {\sin(n+1)\theta\over
{\sin\theta}}-{h\over 2}\, {\sin(n-1)\theta\over {\sin\theta}}
 ,\   x=h\cos\theta.
  \end{equation}

\subsection{Generalized Che\-by\-shev\\ polynomials of the second kind}

Comparing~(\ref{sV}) and~(\ref{mV}), we obtain firstly unified
Che\-by\-shev polynomials of the second kind (for $h=1$ and $h=2$).
Second, these formulas are used to introduce the generalized ones by
declaring the hyperkind $h$ to be any positive integer.

Thus, the generalized Che\-by\-shev polynomials of the 
second kind ${ V}^{(h)}_{n}(x)$ are defined  as\vspace*{-1mm}
\begin{equation}\label{gV}{ V}^{(h)}_{n}(x)={\sin(n+1)\theta\over {\sin\theta}},\quad
x=h\cos\theta ,
  \end{equation}
with $h$ to be a positive integer.\,\,Relation~(\ref{sinsin}) yields
definition~(\ref{gV}) in terms of the  recurrence
relation\vspace*{-1mm}
 \begin{equation}\label{rec-gV} { V}^{(h)}_{n+1}={2\over h}x{ V}^{(h)}_{n}-{ V}^{(h)}_{n-1},~ {
 V}^{(h)}_{0}=1,~
{V}^{(h)}_{1}={2\over h}x.
 \end{equation}

In view of~(\ref{rec-gV}), some  first generalized Che\-by\-shev
polynomials of the second kind are as follows:\vspace*{-1mm}
\begin{equation} \label{gVx}
 \begin{array}{l}
\displaystyle{V}^{(h)}_{0}=1,\ { V}^{(h)}_{1}={2\over h}x,~ {
V}^{(h)}_{2}={4\over h^2}x^{2}-1,
\\[3mm]
\displaystyle{ V}^{(h)}_{3}={8\over h^3}x^{3}-{4\over h}x,~
{V}^{(h)}_{4}={16\over
h^4}x^{4}-{12\over h^2}x^{2}+1, \\[3mm]
\displaystyle { V}^{(h)}_{5}={32\over h^5}x^{5}-{32\over
h^3}x^{3}+{6\over h}x.
 \end{array}\end{equation}
By substituting $h=1$ into~(\ref{gVx}), one has standard
Che\-by\-shev polynomials of the second kind; if $h=2$~-- monic
ones.

 From formula~(\ref{2cos}),
one finds that Che\-by\-shev polynomials $T^{(h)}_{n}(x)$ and
$V^{(h)}_{n}(x)$ are connected by the formula\vspace*{-1mm}
 \begin{equation}\label{con1}
2T^{(h)}_{n}(x)=h\Bigl(\!V^{(h)}_{n}(x)-V^{(h)}_{n-2}(x)\!\Bigr),
\quad  n\ge 0,\end{equation}\vspace*{-5mm}

\noindent where $V^{(h)}_{-1}=0$, $V^{(h)}_{-2}=-1\ $
from~(\ref{rec-gV}).

For example, (generalized) Che\-by\-shev polynomials of the second
kind of the third hyperkind, $(k=2$, $ h=3)$, are:\vspace*{-1mm}
\begin{equation}\label{3V}{ V}^{(3)}_{n}(x)={\sin(n+1)\theta\over {\sin\theta}},\quad
3\cos\theta=x,
  \end{equation}\vspace*{-9mm}
\begin{equation}\label{rec-3V} { V}^{(3)}_{n+1}=\frac{2} {3}x{ V}^{(3)}_{n}-{ V}^{(3)}_{n-1},~ {
V}^{(3)}_{0}=1,~ {V}^{(3)}_{1}={2\over 3}x,
 \end{equation}\vspace*{-9mm}
 \begin{equation}\label{3Vx}
 \begin{array}{l}
\displaystyle{V}^{(3)}_{0}=1,~ { V}^{(3)}_{1}={2\over 3}x,~ {
V}^{(3)}_{2}={4\over 9}x^{2}-1, \\[3mm]
\displaystyle{ V}^{(3)}_{3}={8\over 27}x^{3}-{4\over 3}x,~
{V}^{(3)}_{4}={16\over
81}x^{4}-{12\over 9}x^{2}+1, \\[3mm]
 \displaystyle{ V}^3_{5}={32\over 243}x^{5}-{32\over 27}x^3+2x.
 \end{array} \end{equation}

\section{Vertical Generalization of Che\-by\-shev \\ Polynomials: Expanding Kind \boldmath$k$ to Infinity}

The next direction of a generalization of Che\-by\-shev polynomials
is the vertical generalization.\,\,This means expanding the meaning
of kind $k$ to infinity.\,\,The main idea for the vertical
generalization is the {\it method of equidistant coefficients.}

We now introduce the following notation for generalized
Che\-by\-shev polynomials: $T^{(k,h)}_n(x)$.\,\,They are connected
with the abovementioned polynomials in such way:
$T^{(1,1)}_n(x)=T^{(1)}_n(x)$,  $T^{(1,2)}_n(x)=T^{(2)}_n(x)$,
$T^{(1,h)}_n(x)=T^{(h)}_n(x)$, $T^{(2,1)}_n(x)=V^{(1)}_n(x)$,
$T^{(2,2)}_n(x)=$ $=V^{(2)}_n(x)$,
$T^{(2,h)}_n(x)=V^{(h)}_n(x)$.\,\,Now, we are going to generalize
standard (and monic) Che\-by\-shev polynomials by expanding the
meaning of kind $k$.

\subsection{Equidistant standard\\ Che\-by\-shev polynomials}

The polynomials introduced here can be also called generalized
Che\-by\-shev polynomials of the first hyperkind or equidistant
Che\-by\-shev polynomials of the first hyperkind.

As an example, let us introduce the standard Che\-by\-shev
polynomials of the {\it third kind ${ T}^{(3,1)}_{n}(x)$}, i.e.,
  $(k=$ $=3$, $h=1)$.\,\,Comparing the introductions of $T^{(1,1)}_n(x)$ by~(\ref{rec-sT}) and  $T^{(2,1)}_n(x)$ by~(\ref{rec-sV})  (i.e., $T^{(1)}_n(x)$ and $ V^{(1)}_n(x)$),
  one obtains, as a result of the {natural generalization,} the following definition:
 \begin{equation}\label{rec-s3}\begin{array}{l} {T}^{(3,1)}_{n+1}(x)=2x{T}^{(3,1)}_{n}(x)-{ T}^{(3,1)}_{n-1}(x),\\[2mm]
  { T}^{(3,1)}_{0}=1,~
{ T}^{(3,1)}_{1}=3x. \end{array}
\end{equation} Some examples of ${T}^{(3,1)}_{n}(x)$  are:
 \begin{equation}\label{s3x}
 \begin{array}{l}
{T}^{(3,1)}_{0}=1,~ { T}^{(3,1)}_{1}=3x,~
   {T}^{(3,1)}_{2}=6x^{2}-1, \\[2mm]
{T}^{(3,1)}_{3}=12x^{3}-5x,~
{T}^{(3,1)}_{4}=24x^{4}-16x^{2}+1,\\[2mm]
{T}^{(3,1)}_{5}=48x^{5}-44x^{3}+7x.
 \end{array}\end{equation}
By using~(\ref{sTx}), (\ref{sVx}), and~(\ref{s3x}),
it is easy to verify that
\[{T}^{(3,1)}_{n}-{T}^{(2,1)}_{n}={T}^{(2,1)}_{n}-{T}^{(1,1)}_{n},\]\vspace*{-9mm}
\[
n=1,2,3,4,5,
\]
 which demonstrates the property of equidistancy
of the coefficients.

By analogy, the definition of standard Che\-by\-shev polynomials of
{\it arbitrary kind} ${ T}^{(k,1)}_{n}(x)$ is:
 \begin{equation}\label{rec-sk}
 \begin{array}{l} {T}^{(k,1)}_{n+1}(x)=2x{T}^{(k,1)}_{n}(x)-{ T}^{(k,1)}_{n-1}(x),  \\[2mm]
  { T}^{(k,1)}_{0}=1,\quad
{ T}^{(k,1)}_{1}=kx.\end{array}
\end{equation}
Some first examples of
${ T}^{(k,1)}_{n}(x)$ are:
 \begin{equation}
  \label{skx}
 \begin{array}{l}
{T}^{(k,1)}_{0}=1,~ { T}^{(k,1)}_{1}=kx,\\[1mm]
   {T}^{(k,1)}_{2}=2kx^{2}-1, \\[1mm]
{T}^{(k,1)}_{3}=4kx^{3}-(k+2)x,\\[1mm]
{T}^{(k,1)}_{4}=8kx^{4}-4(k+1)x^{2}+1,\\[1mm]
{T}^{(k,1)}_{5}=16kx^{5}-4(3k+2)x^{3}+(k+4)x.
 \end{array}
 \end{equation}
 The equidistancy of the coefficients in this case is expressed by the fact that the difference
 \[
 {T}^{(k+1,1)}_{n}-{T}^{(k,1)}_{n}
 \]
 is independent of $k$ for any $n$ (due to the linearity of ${T}^{(k,1)}_{n}$ in $k$).

Putting $k=1$ in~(\ref{skx}), one has~(\ref{sTx}), $k=2$ gives~(\ref{sVx}), and $k=3$ -- (\ref{s3x}).

 Taking~(\ref{sT-}) and~(\ref{sV}) in the form
 \begin{equation}
 \label{sT--}\begin{array}{l}\displaystyle T^{(1,1)}_{n}(x)={1\over 2}\, {\sin(n+1)\theta\over {\sin\theta}}-{1\over 2}\, {\sin(n-1)\theta\over {\sin\theta}},\\[3mm]
 \displaystyle T^{(2,1)}_{n}(x)={1}\, {\sin(n+1)\theta\over {\sin\theta}}+{0}\, {\sin(n-1)\theta\over
 {\sin\theta}},\\[3mm]
   x=\cos\theta ,
\end{array}
\end{equation} and using the property of equidistant
coefficiens, one easily finds the formula for equidistant standard
Che\-by\-shev polynomials of the third kind (of the first
hyperkind):\vspace*{-2mm}
 \begin{equation}\label{s3--}
 \begin{array}{l}\displaystyle T^{(3,1)}_{n}(x)={3\over 2}\, {\sin(n+1)\theta\over {\sin\theta}}+{1\over 2}\, {\sin(n-1)\theta\over {\sin\theta}},\\[3mm]
   x=\cos\theta,
\end{array} \end{equation} and of  arbitrary kind of the first hyperkind:
 \begin{equation}\label{sk--}\begin{array}{l}\displaystyle  T^{(k,1)}_{n}(x)={k\over 2} {\sin(n+1)\theta\over {\sin\theta}}\!+\!{{(k-2)}\over 2} {\sin(n-1)
 \theta\over {\sin\theta}}, \\[3mm]
   x=\cos\theta.
\end{array}\!\!\!\!\!\!\!\!\!\!\!\!\!\!\!\!\!\!
\end{equation}
It is easy to verify, by using (\ref{sinsin}), that (\ref{s3--}) and
(\ref{sk--}) satisfy (\ref{rec-s3}) and (\ref{rec-sk}),
correspondingly.

At last, we prove~(\ref{sk--}). Indeed, the general formula has
two equidistant coefficients $\alpha^{(k,1)}$ and $\beta^{(k, 1)}$ and
looks as\vspace*{-3mm}
\begin{equation}
\label{sk---} T^{(k,1)}_{n}(x)=\alpha^{(k,1)}\,
{\sin(n+1)\theta\over {\sin\theta}}+\beta^{(k,1)}\,
{\sin(n-1)\theta\over {\sin\theta}}.
  \end{equation}
Introducing  the difference of neighboring equidistant
coefficients\vspace*{-1mm}
\[
\delta^{(h)}_{1}=\alpha^{(k+1,h)}-\alpha^{(k,h)},\
\delta^{(h)}_{2}=\beta^{(k+1,h)}-\beta^{(k,h)},
\]
which are idependent of $k,$ we have\vspace*{-1mm}
\[
\alpha^{(k,1)}=\alpha^{(1,1)}+\delta^{(1)}_{1}(k-1)=
\alpha^{(1,1)}\!+\!(\alpha^{(2,1)}-\alpha^{(1,1)})\,\times\]\vspace*{-9mm}
\[
\times\,
(k-1)= {1\over 2}+\left(\!1-{1\over 2}\!\right)(k-1)={k\over 2},
\]\vspace*{-7mm}
\[\beta^{(k,1)}\!=\!\beta^{(1,1)}+\delta^{(1)}_{2}(k-1)=\beta^{(1,1)}\!+\!(\beta^{(2,1)}-\beta^{(1,1)})\,
\times
\]\vspace*{-9mm}
\[
\times\, (k-1)={-{1\over 2}}+\left(\!0-\left(\!{-{1\over
2}}\!\right)\right)(k-1)={{(k-2)}\over 2}.
\]

\subsection{Equidistant monic\\ Che\-by\-shev polynomials}

They can be also called generalized Che\-by\-shev polynomials of the
second hyperkind or equidistant Che\-by\-shev polynomials of the
second hyperkind.

To introduce the monic Che\-by\-shev polynomials of the {\it third
kind ${ T}^{(3,2)}_{n}(x)$}, where $(k=3$, $h=2),$ one must compare
the introductions of $T^{(1,2)}_n(x)$ by~(\ref{rec-mT}) and
$T^{(2,2)}_n(x)$ by~(\ref{rec-mV}):\vspace*{-1mm}
   \begin{equation}\label{rec-m3}
 \begin{array}{l} {T}^{(3,2)}_{n+1}(x)=x{T}^{(3,2)}_{n}(x)-{ T}^{(3,2)}_{n-1}(x),
 \\[1mm]
  { T}^{(3,2)}_{0}=0,~
{ T}^{(3,2)}_{1}=x.\end{array}
\end{equation}

\noindent Below, we give some examples of ${
T}^{(3,2)}_{n}(x)$:\vspace*{-1mm}
 \begin{equation}\label{m3x}
\begin{array}{l} {T}^{(3,2)}_{0}=0,~ { T}^{(3,2)}_{1}=x,~
   {T}^{(3,2)}_{2}=x^{2},\\[1mm]
{T}^{(3,2)}_{3}=x^{3}-x,~ {T}^{(3,2)}_{4}=x^{4}-2x^{2},
\\[1mm]
{T}^{(3,2)}_{5}=x^{5}-3x^{3}+x.
 \end{array}\end{equation}

Let us define monic Che\-by\-shev polynomials of  {\it
arbitrary~kind} ${ T}^{(k,2)}_{n}(x)$:\vspace*{-1mm}
 \begin{equation}\label{rec-mk}\begin{array}{l}
 {T}^{(k,2)}_{n+1}(x)=x{T}^{(k,2)}_{n}(x)-{ T}^{(k,2)}_{n-1}(x),
 \\[1mm]
  { T}^{(k,2)}_{0}=3-k,~
{ T}^{(k,2)}_{1}=x.\end{array}
\end{equation}
The first examples of ${ T}^{(k,2)}_{n}(x)$ are as follows:\vspace*{-1mm}
\begin{equation} \label{mkx}
 \begin{array}{l}
{T}^{(k,2)}_{0}=3-k,\
 { T}^{(k,2)}_{1}=x,  \\[1mm]
   {T}^{(k,2)}_{2}=x^{2}+(k-3),\
{T}^{(k,2)}_{3}=x^{3}+(k-4)x,  \\[1mm]
{T}^{(k,2)}_{4}=x^{4}+(k-5)x^{2}-(k-3), \\[1mm]
{T}^{(k,2)}_{5}=x^{5}+(k-6)x^{3}-(2k-7)x.
 \end{array} \end{equation}

 Comparing~(\ref{mT-}) and~(\ref{mV}) in the form\vspace*{-1mm}
 \begin{equation}\label{mT--}\begin{array}{l}
 \displaystyle T^{(1,2)}_{n}(x)={1}\, {\sin(n+1)\theta\over {\sin\theta}}-{1}\,{\sin(n-1)\theta\over
  {\sin\theta}},
  \\[3mm]
\displaystyle T^{(2,2)}_{n}(x)={1}\, {\sin(n+1)\theta\over
{\sin\theta}}+{0}\, {\sin(n-1)\theta\over {\sin\theta}},
   \\[2mm]
 x=2\cos\theta ,
\end{array}   \end{equation} and using  equidistant coefficiens, one obtains
the formula for equidistant monic Che\-by\-shev polynomials of
arbitrary kind:\vspace*{-1mm}
 \begin{equation}\label{mk--}\begin{array}{l}
 \displaystyle T^{(k,2)}_{n}(x)={\sin(n+1)\theta\over {\sin\theta}}+{{(k-2)}}\, {\sin(n-1)\theta\over {\sin\theta}}, \\[2mm]
 x=2\cos\theta.
\end{array}  \end{equation}

\subsection{Equidistant Che\-by\-shev\\ polynomials  of kind \boldmath$k$ of hyperkind \boldmath$h$}

Taking~(\ref{gT-}) and~(\ref{gV}) in the
form\vspace*{-1mm}
 \begin{equation}\label{gT--}\begin{array}{l}
\displaystyle  T^{(1,h)}_{n}(x)
 ={h\over 2}\, {\sin(n+1)\theta\over {\sin\theta}}-{h\over 2}\, {\sin(n-1)\theta\over {\sin\theta}},
   \\[3mm]
\displaystyle T^{(2,h)}_{n}(x)={1}\, {\sin(n+1)\theta\over
{\sin\theta}}+{0}\,{\sin(n-1)\theta\over {\sin\theta}},
 \\[2mm]
  x=h\cos\theta ,
  \end{array} \end{equation}
and using the property of equidistant coefficiens, one easily finds
the formula for equidistant standard Che\-by\-shev polynomials of
arbitrary kind:\vspace*{-1mm}
 \[
  T^{(k,h)}_{n}(x)=\Bigl( {(k-1)-(k-2){h\over 2}}\Bigr) \, {\sin(n+1)\theta\over {\sin\theta}}\,+
  \]\vspace*{-7mm}
\begin{equation}\label{gkh}
+\,\Bigl(\! (k-2){h\over 2}\!\Bigr)\, {\sin(n-1)\theta\over
{\sin\theta}},\quad
    x=h\cos\theta .
 \end{equation}
 Indeed, the general formula has two equidistant
coefficients $\alpha^{(k,h)}$ and $\beta^{(k,h)}$  and looks
as\vspace*{-1mm}
\begin{equation}\label{sk---} T^{(k,h)}_{n}(x)=\alpha^{(k,h)}\,
{\sin(n+1)\theta\over {\sin\theta}} +\beta^{(k,h)}\,
{\sin(n-1)\theta\over {\sin\theta}},
  \end{equation}\vspace*{-7mm}

\noindent where\vspace*{-1mm}
\[\alpha^{(k,h)}= \alpha^{(1,h)}+\delta^{(h)}_{1}(k-1)=\]\vspace*{-9mm}
\[ =\alpha^{(1,h)}+(\alpha^{(2,h)}-\alpha^{(1,h)}) (k-1)=
(k-1)-(k-2){h\over 2},\]\vspace*{-7mm}
\[\beta^{(k,h)}=\beta^{(1,h)}+\delta^{(h)}_{2}(k-1)=\]\vspace*{-9mm}
\[\beta^{(1,h)}+(\beta^{(2,h)}-\beta^{(1,h)})(k-1)=
 (k-2){h\over 2}.  \]\vspace*{-5mm}

 From~(\ref{rec-gT}) and~(\ref{gkh}), we have\vspace*{-1mm}
 \begin{equation}\label{rec-Tkh}
 \begin{array}{l}
 \displaystyle{T}^{(k,h)}_{n+1}(x)={2\over h}x{T}^{(k,h)}_{n}(x)-{ T}^{(k,h)}_{n-1}(x),
   \\[2mm]
 T^{(k,h)}_{0}(x)= {(k-1)-(k-2){h}}=A ,
  \\[1mm]
 \displaystyle T^{(k,h)}_{1}(x)=\bigl( (k-1){2\over h}-(k-2)\bigr)x=Bx,
  \end{array}
  \end{equation}\vspace*{-3mm}

 \noindent
which can be the definition of $T^{(k,h)}_{n}(x)$.

At last, from~(\ref{rec-Tkh}), we have some first examples of ${
T}^{(k,h)}_{n}(x)$:\vspace*{-3mm}
 \begin{equation} \label{Tkhx}
 \begin{array}{l}
\displaystyle {T}^{(k,h)}_{0}=A,\ { T}^{(k,h)}_{1}=Bx,\
   {T}^{(k,h)}_{2}={2\over h}Bx^{2}-A,
  \\[3mm]
\displaystyle {T}^{(k,h)}_{3}={4\over h^2}Bx^{3}-\left(\!{2\over
h}A+B\!\right)\!x,
  \\[5mm]
\displaystyle {T}^{(k,h)}_{4}={8\over h^3}Bx^{4}-\left(\!{4\over
h^2}A+{4\over h}B\!\right)\!x^2+A,
  \\[5mm]
\displaystyle {T}^{(k,h)}_{5}={16\over h^4}Bx^{5}-\left(\!{8\over
h^3}A+{12\over h^2}B\!\right)\!x^3+\left(\!{4\over
h}A+B\!\right)\!h.
 \end{array}\!\!\!\!\!\!\!\!\!\!\!\!\!\!\!\!\!\!\!\!\!\!\!\!\!\!\!\!
 \end{equation}

From~(\ref{gkh}), we obtain\vspace*{-1mm}
\begin{equation}\label{gkh-}
T^{(k,h)}_{n}=(k-1)T^{(2,h)}_{n}-(k-2)T^{(1,h)}_{n},\end{equation}
and\vspace*{-1mm}
\[
T^{(k,h)}_{n}(x)=\bigl( {(k-1)-(k-2){h\over 2}}\bigr)
{T_{n}^{(2,h)}}\,+
\]\vspace*{-7mm}
\begin{equation}\label{gkh--}
 +\,\biggl(\! (k-2){h\over 2}\!\biggr) {T_{n-2}^{(2,h)}},\quad
    x=h\cos\theta .\end{equation}

\section{Che\-by\-shev Polynomials\\ and Alexander Knot Invariants}

The Alexander polynomials $\Delta(q)$ are defined by the Alexander
skein relation for knots and links~\cite{Al}, (see notations
in~\cite{Pa-agg}):\vspace*{-1mm}
 \begin{equation}\label{alex-skein}
\Delta_+(q)-\Delta_-(q)=(q^{1\over2}-q^{-{1\over2}})\Delta_{O}(q)
,\quad \Delta_{\rm unknot}=1.
 \end{equation}
 The Alexander polynomial invariants for torus knots
$T(n,l)$ are given by the formula~\cite{Ro,Li}\vspace*{-1mm}
\begin{equation}\label{deltal}
 {\Delta}_{n,l}(q)=
{{(\ts q^{nl\over 2}-q^{-{nl\over 2}})(q^{1\over 2}-q^{-{1\over
2}})} \over{(\ts q^{n\over 2}-q^{-{n\over 2}})(q^{l\over
2}-q^{-{l\over 2}})}},
\end{equation}
 where $n$ and $l$ are coprime positive integers.
For $l=$ $=2$, formula (\ref{deltal}) gives the expression for torus
knots $T(n,2)$:\vspace*{-3mm}
 \begin{equation}\label{deltal2}
{\Delta}^{(K)}_{n,2}(q)= {{q^{n\over 2}+q^{-{n\over 2}}} \over{q^{1\over
2}+q^{-{1\over 2}}}},\quad n=2m-1, 
 \end{equation}
where $m$ are positive integers: $m=1,2,3,...\,$.\,\,Here, $(K)$
stands for a ``knot'' to be distinguished from a ``link''
$(L)$.\,\,Putting $q={e}^{2i\theta}$ in~(\ref{deltal2}), we
get\vspace*{-1mm}
\begin{equation}\label{deltal2q}
{\Delta}^{(K)}_{n,2}(q)=
 {\cos n\theta\over\cos\theta}, \quad 2\cos\theta=q^{1\over 2}+q^{-{1\over 2}}.
\end{equation}
It easy to see that relations (\ref{gT}) and
(\ref{deltal2q}) yield\vspace*{-1mm}
\begin{equation}\label{deltal2theta}
{T^{(1,h)}_{n}(x)\over x}= {\cos
n\theta\over\cos\theta}={\Delta}_{n,2}^{(K)}(q), \quad 2\cos\theta
=q^{1\over 2}+q^{-{1\over 2}}
\end{equation}
for odd $n$.\,\,This formula says firstly that the Alexander
polynomials ${\Delta}_{n,2}^{(K)}(q)$ are invariants (not depending
on $h$) of all generalized Che\-by\-shev polynomials of the first
kind.\,\,Second, it gives the recipe for obtaining the corresponding
Che\-by\-shev polynomials (for odd $n$) from the Alexander
polynomials ${\Delta}_{n,2}^{(K)}(q)$:\vspace*{-1mm}
\begin{equation}\label{t1h}{T^{(1,h)}_{n}(x)}=
x{\Delta}_{n,2}^{(K)}(q),  \quad  x={h\over 2}(q^{1\over
2}+q^{-{1\over 2}}).\end{equation}

The corresponding formula for torus links $L(n,2)$ (in the case of
even $n$)~\cite{Pa-agg} are:\vspace*{-1mm}
\begin{equation}\label{deltal2-l}
{\Delta}^{(L)}_{n,2}(q)={{q^{n\over 2}-q^{-{n\over 2}}}
\over{q^{1\over 2}+q^{-{1\over 2}}}} ,\quad n=2m.
 \end{equation}

It can be shown that, for odd $n,$%
\begin{equation}\label{t2h}{T^{(2,h)}_{n}(x)}=
{2x\over h}{{{\Delta}_{n+1,2}^{(L)}(q)}\over(q^{1\over
2}-q^{-{1\over 2}})} ,
 \quad  x={h\over 2}(q^{1\over 2}+q^{-{1\over 2}}).\end{equation}

 At last, we give some Alexander polynomials, which can be calculated from~(\ref{deltal2}) and (\ref{deltal2-l})
 and can be used for the direct verification of~(\ref{t1h}) and (\ref{t2h}):\vspace*{-1mm}
\begin{equation}\label{alex2}
 \begin{array}{l}
 {\Delta}_{1,2}^{(K)}(q)=1,\quad { \Delta}_{2,2}^{(L)}(q)=q^{1\over 2}-q^{-{1\over 2}},
  \\[1mm]
 { \Delta}_{3,2}^{(K)}(q)=q-1+q^{-{1}},
   \\[1mm]
  { \Delta}_{4,2}^{(L)}(q)=q^{3\over 2}-q^{1\over 2}+q^{-{1\over 2}}-q^{-{3\over 2}},
 \\[1mm]
  { \Delta}_{5,2}^{(K)}(q)= q^{2}-q+1-q^{-1}+q^{-{2}}.
 \end{array}\end{equation}

\section{Concluding Remarks}

 We have proposed a rather simple generalization of Che\-by\-shev polynomials, which follows immediately
 from the attempt to unify standard and monic Che\-by\-shev polynomials of both kinds
 with the help of one formula.\,\,Is is believed that these polynomials are connected
 with some knot and link polynomial invariants.\,\,Some connections of them with
 Alexander polynomial invariants are presented here.\,\,We believe that
 a deeper investigation in this direction will bring interesting results.

\vskip3mm {\it The present work was partially supported by the
National Academy of Sciences of Ukraine (project No.\,0117U000238).}


\vspace*{-5mm}
\rezume{%
A.M.\,Павлюк} {УЗАГАЛЬНЕНІ ЕКВІДИСТАНТНІ ПОЛІНОМИ \\ЧЕБИШОВА І
ВУЗЛОВІ ІНВАРІАНТИ АЛЕКСАНДЕРА}
 {Вводяться узагальнені еквідистантні поліноми Чебишова  $T^{(k,h)}$ роду $k$ із гіперроду $h,$
де $k,h$ -- додатні цілі числа. Вони отримані шляхом узагальнення
стандартних поліномів Чебишова першого і другого родів та монічних
поліномів Чебишова першого і другого родів. Це узагальнення
виконується в двох напрямках. Горизонтальне узагальнення
здійснюється шляхом введення поняття гіперроду $h$, і розширення
множини його значень до нескінченності.  Вертикальне узагальнення
передбачає розширення множини значень роду $k$ до нескінченності
методом еквідистантних коефіцієнтів. Досліджено деякі зв'язки
введених поліномів з вузловими інваріантами Александера.}

\end{document}